\documentclass[journal,twoside,web]{ieeecolor}
\usepackage{jsen}
\usepackage{cite}
\usepackage{amsmath,amssymb,amsfonts}
\usepackage{algorithm}
\usepackage{algorithmic}
\usepackage{array}
\usepackage[caption=false,font=normalsize,labelfont=sf,textfont=sf]{subfig}
\usepackage{textcomp}
\usepackage{stfloats}
\usepackage{url}
\usepackage{verbatim}
\usepackage{graphicx}
\usepackage{subcaption}
\usepackage{wrapfig}
\usepackage{cuted}
\usepackage{caption}
\usepackage{xcolor} 
\usepackage{hyperref}
\def\BibTeX{{\rm B\kern-.05em{\sc i\kern-.025em b}\kern-.08em
    T\kern-.1667em\lower.7ex\hbox{E}\kern-.125emX}}

\definecolor{abstractbg}{rgb}{0.89804,0.94510,0.83137}
\begin{document}

\title{VIRTUS-FPP: Virtual Sensor Modeling for Fringe Projection Profilometry in NVIDIA Isaac Sim}

\author{Adam Haroon$^{1\$}$, Anush Lakshman$^{1\$}$, Badrinath Balasubramaniam$^{2}$, Beiwen Li$^{2*}$
\thanks{\copyright\,2026 IEEE. Personal use of this material is permitted. Permission from IEEE must be obtained for all other uses, in any current or future media, including reprinting/republishing this material for advertising or promotional purposes, creating new collective works, for resale or redistribution to servers or lists, or reuse of any copyrighted component of this work in other works. Digital Object Identifier: 10.1109/JSEN.2026.3698278}
\thanks{This article has been accepted for publication in IEEE Sensors Journal. Manuscript received xx, 2026; revised xx, 2026. \textit{(Corresponding author: Beiwen Li)}}
\thanks{$^{1}$Adam Haroon and Anush Lakshman are with the Department of
        Mechanical Engineering, Iowa State University, Ames, IA 50012, USA
        {\tt\small anushlak@iastate.edu; aharoon@iastate.edu}}%
\thanks{$^{2}$Badrinath Balasubramaniam and Beiwen Li are with the College of Engineering,
        University of Georgia, Athens, GA 30602, USA
        {\tt\small bb2@uga.edu; beiwen.li@uga.edu}}%
\thanks{$^{\$}$These authors contributed equally.}
}

\IEEEtitleabstractindextext{%
\fcolorbox{abstractbg}{abstractbg}{%
\begin{minipage}{\textwidth}%
\begin{wrapfigure}[12]{r}{3in}%
\includegraphics[width=3in]{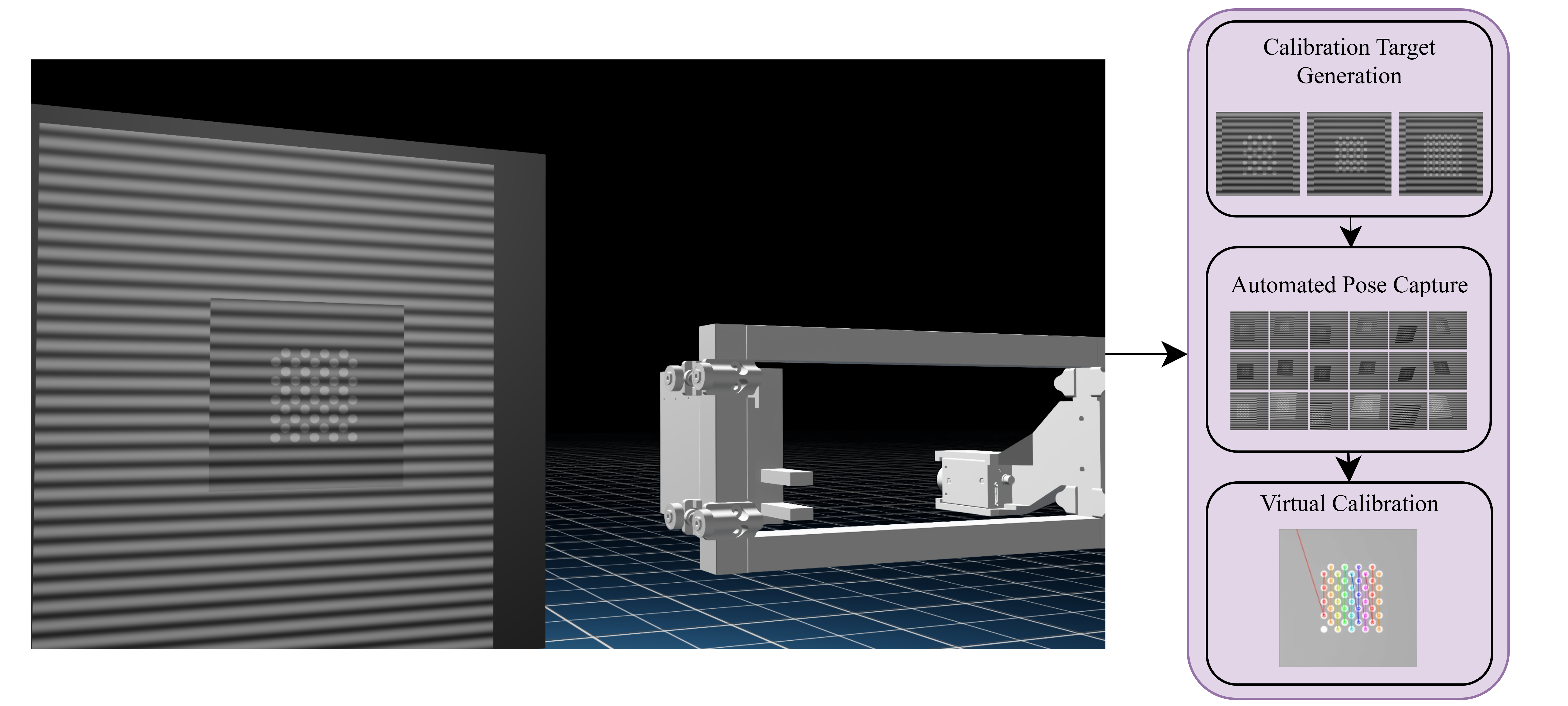}%
\end{wrapfigure}%
\begin{abstract}
Fringe projection profilometry (FPP) is a high-precision structured-light sensing technique for 3D surface reconstruction, yet its practical deployment is often constrained by complex calibration procedures, sensitivity to environmental conditions, and the high cost of physical experimentation. At the same time, robotics research increasingly relies on simulation platforms such as NVIDIA Isaac Sim for scalable development and validation, but accurate virtual representations of optical metrology sensors such as FPP are not currently available. In this work, we present VIRTUS-FPP, the first end-to-end virtual sensor modeling framework for fringe projection profilometry implemented in NVIDIA Isaac Sim, enabling physically grounded simulation of the complete FPP pipeline, including structured light projection, image formation, calibration, and 3D reconstruction, without dependence on pre-calibrated physical systems. The framework leverages an inverse camera model for projector representation, ensuring geometric and photometric fidelity consistent with structured-light principles. By bridging optical metrology and robotics simulation, VIRTUS-FPP enables high-fidelity synthetic data generation, systematic evaluation of sensing pipelines, and digital twin replication of real-world FPP systems. Experimental results demonstrate sub-millimeter reconstruction accuracy and strong correspondence between simulated and physical measurements, highlighting the framework's effectiveness and its potential to advance perception-driven robotics, simulation-to-reality transfer, and scalable optical sensor design.
\end{abstract}

\end{minipage}}}

\maketitle

\begin{figure*}[htbp]
    \centering
    \includegraphics[width=\textwidth]{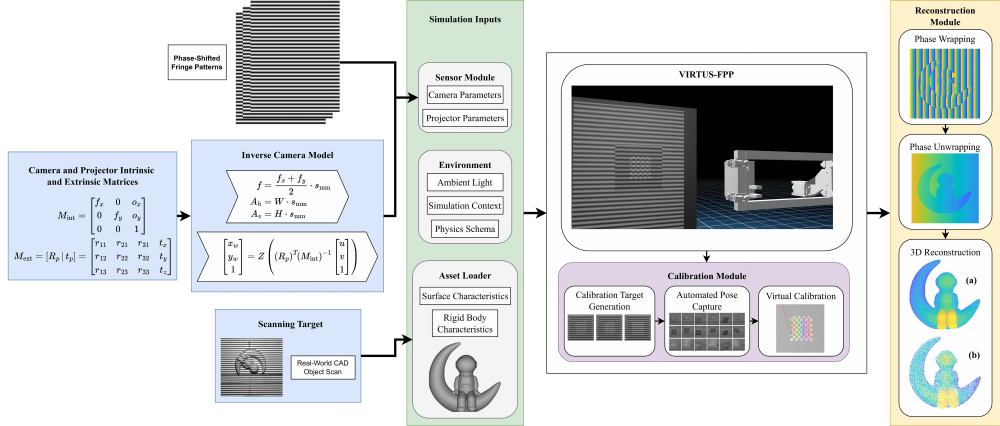}
    \caption{VIRTUS-FPP system architecture and processing pipeline. The framework integrates
    camera and projector intrinsic/extrinsic parameters, inverse camera modeling, and NVIDIA
    Isaac Sim's RTX ray-traced rendering to enable end-to-end virtual FPP.
    The pipeline encompasses fringe pattern generation, virtual calibration target
    positioning, automated pose capture, fringe image capture, phase wrapping/unwrapping,
    and virtual 3D reconstruction (a) with ground truth validation (b).}
    \label{fig:pipe_virtusfpp}
\end{figure*}

\section{Introduction}\label{intro}

\IEEEPARstart{F}{ringe} Projection Profilometry (FPP) is a structured-light 3D imaging
sensor technology offering high surface measurement accuracies through
camera-projector triangulation~\cite{zhang2016high,geng2011structured}.
Despite broad applicability in manufacturing monitoring~\cite{zhang2023machine,qian2021high},
robotic manipulation~\cite{suresh2021high,bai2024close}, corrosion
analysis~\cite{lakshman2024corrosion}, and biomedical
imaging~\cite{zhao2020shortwave,genovese2006whole}, FPP faces fundamental deployment
barriers: calibration is time-consuming and sensitive to environmental
disturbances~\cite{feng2021calibration,zhang2018absolute}, while complex material properties
and ambient lighting degrade reconstruction
quality~\cite{zhang2010recent,feng2018high,wei2014experimental}.
Systematic characterization of these effects requires extensive physical experimentation that
is costly, slow, and difficult to scale.

Virtual sensor modeling and digital twin methodologies address these limitations by enabling
controlled, repeatable FPP sensor evaluation without physical
hardware~\cite{wright2024dtwinmetrology,vlaeyen2021dtwinoptical}.
Prior work includes: graphics-based FPP digital twins in Blender~\cite{zheng2020fringe}; Unity-based FPP simulators with GUI-adjustable
camera parameters~\cite{ueda2021fringe}; ray-tracing simulators modeling perspective
distortion and lens effects~\cite{zhang2023measurement}; and projection error decoupling
methods to address geometric distortion and grayscale
inconsistency~\cite{tang2025projection}.
However, all existing approaches depend on pre-calibrated physical systems by transferring
real-world calibration matrices into simulation, propagating physical measurement errors.
Moreover, Blender, Unity, and MATLAB lack native support for parallel data collection 
and robotics integration, limiting their suitability for large-scale, high-throughput 
virtual sensor modeling~\cite{blender2024cycles, unity2024hdrp}.

These limitations become particularly acute for machine learning applications, where large,
annotated datasets with exact ground truth are essential for training and benchmarking
models for phase quality estimation, surface reconstruction, and defect
detection~\cite{zuo2022deep}.
Machine learning architectures designed for fringe pattern analysis demonstrate significant
improvements in phase unwrapping, denoising, and reconstruction quality when trained on
synthetic datasets~\cite{zheng2020fringe,wang2022deepspatialphase}.
The effectiveness of these approaches critically depends on the fidelity of the simulation
framework, motivating accurate virtual sensor modeling with verified metric consistency.

We propose VIRTUS-FPP: the first fully end-to-end virtual FPP sensor modeling framework,
built on NVIDIA Isaac Sim~\cite{isaacsim_docs}, providing projective geometric calibration,
RTX ray-traced light transport, and calibrated metric consistency across the entire imaging
pipeline from calibration to 3D reconstruction without dependence on pre-calibrated physical hardware, as shown in Fig.~\ref{fig:pipe_virtusfpp}.
Table~\ref{tab:prior_work} summarizes how VIRTUS-FPP compares to existing virtual FPP
approaches across key capabilities.
Our primary contributions are:
\begin{enumerate}
    \item The first end-to-end virtual FPP sensor model with complete virtual calibration,
          eliminating dependence on pre-calibrated physical systems.
    \item An inverse camera model for accurate virtual projector modeling with projective
          geometric and photometric fidelity, enabling geometrically consistent simulation
          through RTX-accelerated ray tracing.
    \item High-throughput synthetic data generation: 16\,848       fringe images in 1.5 hours with parallel data            collection capabilities.
    \item A validated digital twin of a calibrated real-world FPP system demonstrating
          high-fidelity real-to-virtual correspondence.
\end{enumerate}

\begin{table}[!t]
\caption{Comparison of Virtual FPP Sensor Modeling Approaches}
\label{tab:prior_work}
\centering
\setlength{\tabcolsep}{3.5pt}
\renewcommand{\arraystretch}{1.15}
\begin{tabular}{lcccc}
\hline\hline
\textbf{Feature} & \textbf{Zheng}~\cite{zheng2020fringe} & \textbf{Ueda}~\cite{ueda2021fringe} & \textbf{Zhang}~\cite{zhang2023measurement} & \textbf{VIRTUS} \\
 & \textbf{(2020)} & \textbf{(2021)} & \textbf{(2023)} & \textbf{(ours)} \\
\hline
Platform             & Blender   & Unity     & MATLAB    & Isaac Sim   \\
Photorealistic   & \checkmark & $\circ$  & --        & \checkmark  \\
Virtual calibration  & --        & --        & --        & \checkmark  \\
Projector model      & Empirical & Empirical & Geometric & Geometric   \\
Digital twin valid.  & $\circ$   & $\circ$  & --        & \checkmark  \\
Robotics native      & --        & --        & --        & \checkmark  \\
\hline
\multicolumn{5}{l}{\footnotesize \checkmark\,=\,supported;\ $\circ$\,=\,partial;\ --\,=\,not supported;\ N/R\,=\,not reported.}\\
\hline\hline
\end{tabular}
\end{table}

The remainder of this paper is organized as follows.
Section~\ref{methodology} establishes the theoretical foundations of FPP and the
simulation capabilities of NVIDIA Isaac Sim.
Section~\ref{experiments} details the VIRTUS-FPP system architecture and implementation.
Section~\ref{calib_recon} describes the virtual calibration procedure.
Section~\ref{validation} presents experimental validation including reconstruction accuracy,
digital twin fidelity, and 3D reconstruction comparisons.
Section~\ref{discussion} analyzes the framework's advantages and limitations, followed by
conclusions in Section~\ref{conclusion}.

\section{FPP Principles and Simulation Platform} \label{methodology}

\subsection{Fringe Projection Profilometry} \label{fpp}

\begin{figure}[ht]
    \centering
    \includegraphics[scale = 0.64]{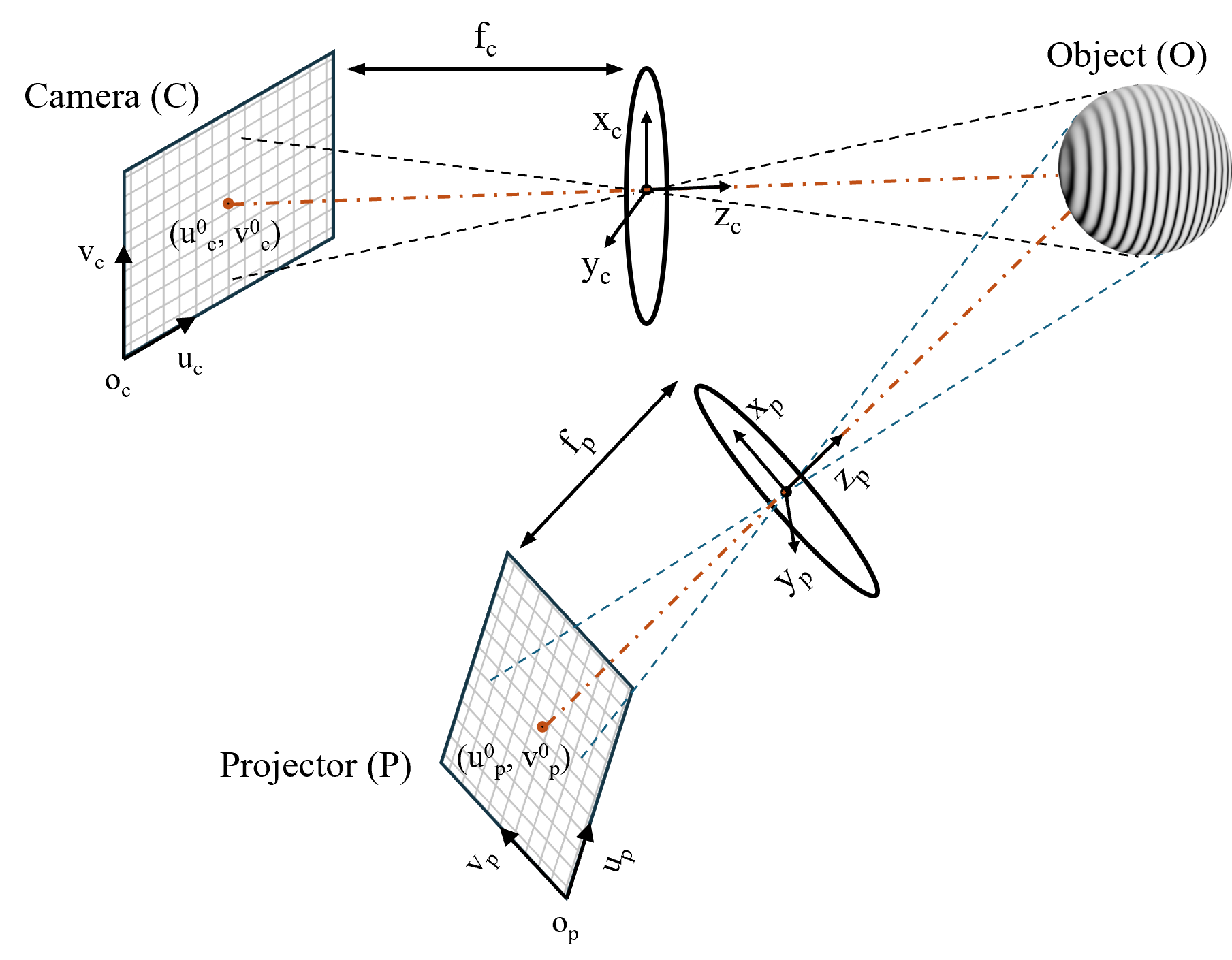}
    \caption{Geometric triangulation principles of FPP: camera-projector configuration and
    structured light projection onto a target object.}
    \label{fig:princi_fpp}
\end{figure}

FPP employs a camera-projector pair (Fig.~\ref{fig:princi_fpp}) where the projector
displays sinusoidal fringe patterns that deform across an object's surface, encoding
geometry as phase information decoded through three steps.

\textbf{Phase Wrapping:} The intensity of the $n^{\text{th}}$ fringe image is:
\begin{equation} \label{phase_shift_math}
    I_n(x,y) = I'(x,y) + I''(x,y)\cos\!\left(\phi(x,y) + \tfrac{2\pi n}{N}\right),
\end{equation}
where $N$ is the number of phase steps, $I'$ is the background intensity, $I''$ is the
modulation amplitude, and $\phi$ is the phase.
We use an 18-step method ($N=18$), applied independently to vertical and horizontal fringe sets, yielding the wrapped phase map:
\begin{equation} \label{pwrap}
    \phi(x,y) = -\tan^{-1}\!\left(\frac{\sum_{n=1}^{N} I_n \sin \delta_n}
                                       {\sum_{n=1}^{N} I_n \cos \delta_n}\right), \quad
    \delta_n = \tfrac{2\pi n}{N}.
\end{equation}
The 18-step method was selected for its high noise immunity, as the large number of phase
steps allows robust phase extraction through averaging across captures, enabling sub-millimeter
accuracy even under moderate scene noise~\cite{zhang2016high}.

\textbf{Phase Unwrapping:} Temporal phase unwrapping with Gray coding resolves $2\pi$
discontinuities:
\begin{equation} \label{puwrap}
    \Phi(x,y) = \phi(x,y) + 2\pi k(x,y).
\end{equation}
We use 7 gray-coded binary patterns projected alongside the fringe patterns to assign an integer
fringe order $k(x,y)$ to each pixel, enabling unambiguous absolute phase recovery without
spatial filtering or phase continuity assumptions. Equations~\ref{pwrap} and~\ref{puwrap} are applied independently to the vertical and horizontal fringe captures, producing two absolute phase maps $\Phi_v(x_c, y_c)$ and $\Phi_h(x_c, y_c)$ at each camera pixel $(x_c, y_c)$.

\textbf{Phase-to-Projector Correspondence:} The two absolute phase maps map directly to projector pixel coordinates, establishing pixel-level correspondence between the camera and projector image planes:
\begin{equation}\label{eq:phase2pix}
u^p = \frac{\Phi_v(x_c, y_c)}{2\pi}\, T_v, \qquad
v^p = \frac{\Phi_h(x_c, y_c)}{2\pi}\, T_h,
\end{equation}
where $T_v$ and $T_h$ are the projector pixel periods (projector pixels per $2\pi$ of phase) of the vertical and horizontal fringe patterns, respectively. This per-pixel $(u^p, v^p)$ correspondence is the geometric input to the triangulation step below.

\textbf{3D Reconstruction:} Given the camera pixel $(u^c, v^c)$ and the matching projector pixel $(u^p, v^p)$ recovered from Eq.~\ref{eq:phase2pix}, camera and projector are modeled as pinhole devices with
transformation matrices $M^c = A^c[R^c|t^c]$ and $M^p = A^p[R^p|t^p]$.
A 3D point $[x,y,z]^T$ relates to image coordinates via:
\begin{equation}
    s^c \!\begin{bmatrix} u^c \\ v^c \\ 1 \end{bmatrix} = M^c \!\begin{bmatrix} x \\ y \\ z \\ 1 \end{bmatrix}, \quad
    s^p \!\begin{bmatrix} u^p \\ v^p \\ 1 \end{bmatrix} = M^p \!\begin{bmatrix} x \\ y \\ z \\ 1 \end{bmatrix}.
\end{equation}
With calibrated $M^c$ and $M^p$, 3D coordinates are recovered by solving
$[x,y,z]^T = \mathbf{A}^{-1}\mathbf{b}$ where $\mathbf{A}$ and $\mathbf{b}$ encode the
respective matrix elements~\cite{zhang2010recent}.

\subsection{NVIDIA Isaac Sim} \label{isaac}

NVIDIA Isaac Sim is a robotics simulation platform built on NVIDIA Omniverse, designed
for developing, testing, and validating AI-based perception systems in high-fidelity
virtual environments~\cite{isaacsim_docs}.
The platform provides programmatic Python API control alongside the OmniGraph visual
interface, enabling fully automated calibration, scanning, and reconstruction pipelines
without manual intervention.

Scene construction is built on Universal Scene Description (USD), Pixar's open-source
framework for describing and composing complex 3D scenes~\cite{usd_docs}.
USD's hierarchical, component-based architecture allows precise definition of camera and
projector geometry, material assignments, and relative spatial relationships, which is
essential for accurately replicating real FPP system configurations in simulation.

Rendering is powered by NVIDIA RTX technology through OptiX ray tracing~\cite{parker2010optix},
providing the light transport simulation essential for fringe pattern
fidelity~\cite{bai2024close,haroon2024autonomous}.
The RTX renderer supports path tracing for global illumination, surface material modeling
via the Material Definition Language (MDL), multi-bounce light transport, and ambient
occlusion.
These effects directly govern fringe pattern deformation on object surfaces, including
reflectivity variations and specular highlights that degrade reconstruction quality in
real systems~\cite{feng2018high,wei2014experimental}.

To clarify the simulation's optical scope, we distinguish three modeling categories. \textit{Geometric optics} is modeled explicitly: ray transport, perspective projection through the pinhole camera/projector model, and surface intersection/occlusion. \textit{Radiometric rendering} is captured through MDL-based surface BRDFs, multi-bounce light transport, global illumination, ambient occlusion, and per-pixel intensity integration via the RTX path tracer. The following effects are \textit{not} explicitly modeled: wave-optical phenomena (coherence, diffraction, interference), projector gamma nonlinearity, lens distortion (radial/tangential), and sensor noise (shot, read). The omission of projector gamma may bias the recovered phase, particularly for low fringe counts, while uncorrected lens distortion can degrade edge accuracy and sensor noise broadens phase variance. These approximations are consistent with the scope of prior virtual FPP work~\cite{zheng2020fringe,ueda2021fringe,zhang2023measurement} and are revisited in Section~\ref{discussion}.

Synthetic data generation is handled by NVIDIA Replicator~\cite{replicator_docs}, which
supports domain randomization across lighting conditions, material properties, and geometric
configurations, as well as automated ground truth generation.
This pipeline enables high-throughput acquisition under controlled, repeatable conditions
without any of the environmental sensitivity that limits physical FPP
experimentation~\cite{feng2021calibration}.

\section{VIRTUS-FPP: System Architecture} \label{experiments}

\subsection{Virtual Camera-Projector System}

We utilize Isaac Sim's pre-defined pinhole camera with an RGB render product for data
acquisition.
The projector is simulated by a rectangular light source (\texttt{UsdLux}) that projects
fringe pattern textures via the \texttt{isProjector} parameter. We choose the rectangular light source as it is the only Isaac Sim light primitive that supports texture projection. Through the inverse camera model 
(Eq.~\ref{eq:p2wfin}), we validate that this light source behaves identically to a pinhole 
projector, with projected pattern dimensions scaling linearly with distance. This validation methodology is discussed in further detail in Section~\ref{Digital Twin Validation} (Fig.~\ref{fig:linearity_viz}).

\begin{figure}[ht]
    \centering
    \includegraphics[width=1\linewidth]{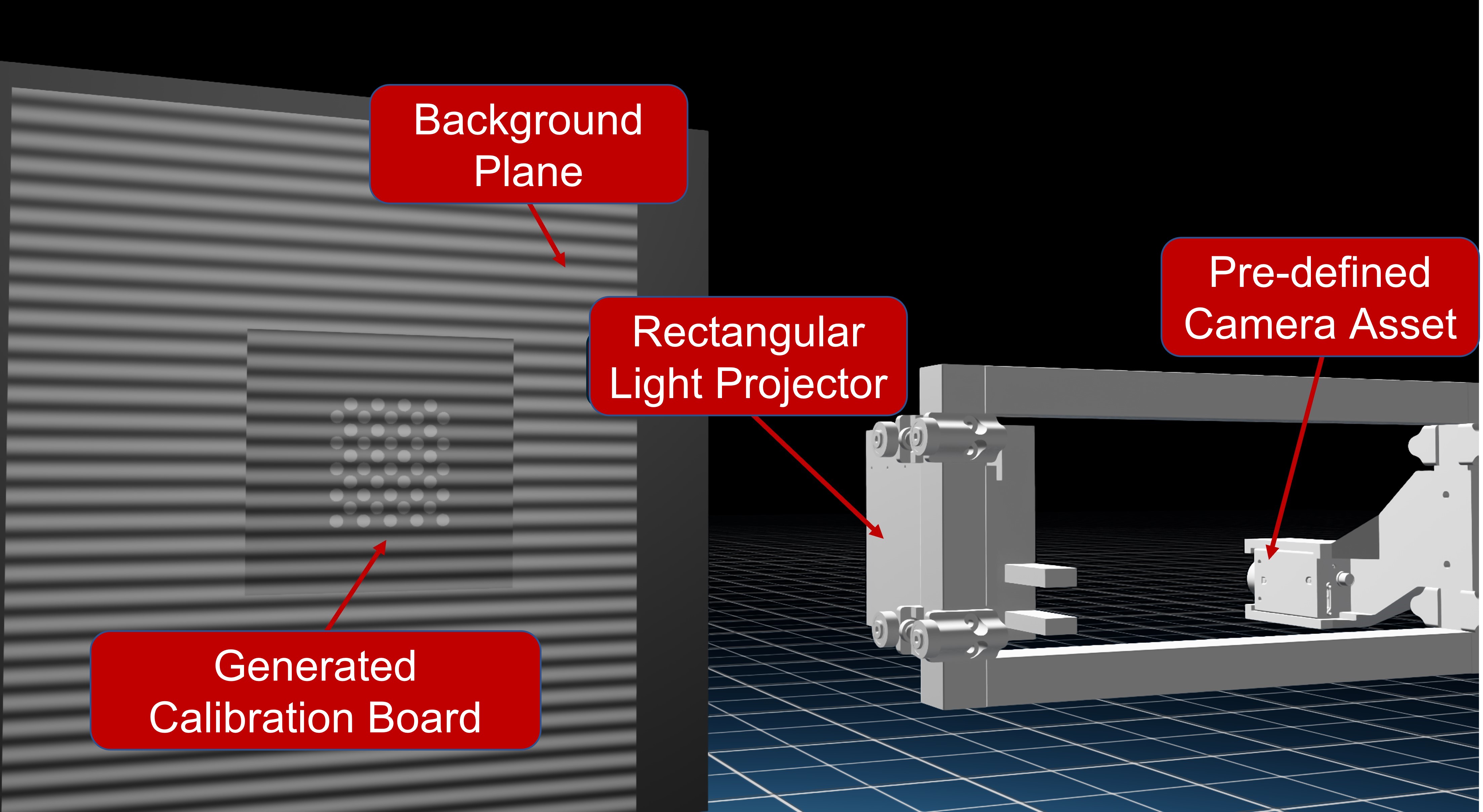}
    \caption{Virtual camera-projector calibration setup with pinhole camera asset,
    rectangular light source projector, generated calibration board, and background plane.}
    \label{fig:simfpp}
\end{figure}

Because the OmniGraph interface cannot dynamically update the texture file we pass to the projector during simulation,
we developed a custom Python extension with three components: scene initialization
(\texttt{setup\_scene}), fringe texture loading and acquisition callbacks
(\texttt{setup\_post\_load}), and episode reset (\texttt{setup\_post\_reset}).
The projector is positioned 0.1 m below and 0.125 m to the left of the camera.
All scanning objects use material properties of 0.95 roughness, 0.15 specular, and 0.95
ambient occlusion-to-diffuse ratio to simulate typical matte surface reflectance
conditions~\cite{mi13101607,ma16155443,wei2014experimental}.
Table~\ref{tab:cam_proj_params} summarizes both asset parameters; the setup is shown in
Figure~\ref{fig:simfpp}.

\begin{table}
\caption{Parameters for Simulated Camera and Projector Assets}
\label{tab:cam_proj_params}
\centering
\begin{tabular}{lc}
\hline\hline
\multicolumn{2}{c}{\textbf{Camera Parameters}} \\
\hline
Parameter & Value \\
\hline
Focal Length          & 50 cm \\
Horizontal Aperture   & 20.9995 cm \\
Vertical Aperture     & 15.2908 cm \\
Resolution            & 960 $\times$ 960 \\
\hline
\multicolumn{2}{c}{\textbf{Projector Parameters}} \\
\hline
Parameter & Value \\
\hline
Intensity             & 40 nits \\
Height                & 0.625 m \\
Width                 & 0.5 m \\
Texture File          & Fringe pattern \\
Projector Light Type  & True \\
\hline\hline
\end{tabular}
\end{table}

\subsection{Synthetic Data Acquisition}

An \texttt{update\_texture\_callback} function sequentially loads each fringe pattern into
the light source's \texttt{texture:file} attribute and captures a grayscale frame after a
brief rendering delay. A complete calibration episode (936 captures across 18 poses) completes in just over
5 minutes through the GPU-accelerated pipeline, averaging 16\,848 fringe image captures
per 1.5 hours.

\section{Virtual Calibration and Reconstruction} \label{calib_recon}

\subsection{Calibration Board Generation}

We procedurally generate asymmetric circular calibration boards as RGB pattern images from
configurable row, column, circle diameter ($D_{\rm circle}$), and circle center distance
($D_{\rm centers}$) parameters.
Each pattern is applied as a diffuse albedo texture map to a plane mesh in the USD scene
using the MDL-based material system.
Because the texture is uniformly scaled to fit the plane geometry, the original image-space
dimensions of the pattern circles are transformed by the plane's aspect ratio and extent.
To recover the true circle dimensions as they exist in the simulation (and
therefore supply accurate calibration geometry to the calibration algorithm), we derive
a closed-form scaling factor from the ratio of the plane dimensions to the pattern's
nominal extent.
First, the pattern dimensions in meters are computed from the image-space parameters as:

\begin{equation}
W_{\rm pat} = \!\left(2L_{\rm b} + \tfrac{(C-1)D_{\rm centers}}{2} + D_{\rm circle}\right)k
\end{equation}
\begin{equation}
H_{\rm pat} = \!\left(2L_{\rm b} + (R-1)D_{\rm centers} + D_{\rm circle}\right)k
\end{equation}

\noindent where $L_{\rm b}$ is the border length, $C$ and $R$ are the number of columns
and rows, and $k = 0.001$ (mm$\,\to\,$m).
A uniform scaling factor is then computed as:

\begin{equation}
S = \min\!\left(\frac{W_{\rm plane}}{W_{\rm pat}},\;\frac{H_{\rm plane}}{H_{\rm pat}}\right)
\end{equation}

\noindent and the true in-simulation circle diameter and center spacing are:

\begin{align}
D_{\rm circle}^{\rm sim} &= D_{\rm circle}\cdot k \cdot S \\
D_{\rm centers}^{\rm sim} &= D_{\rm centers}\cdot k \cdot S
\end{align}

\begin{figure}[ht]
    \centering
    \includegraphics[width=1\linewidth]{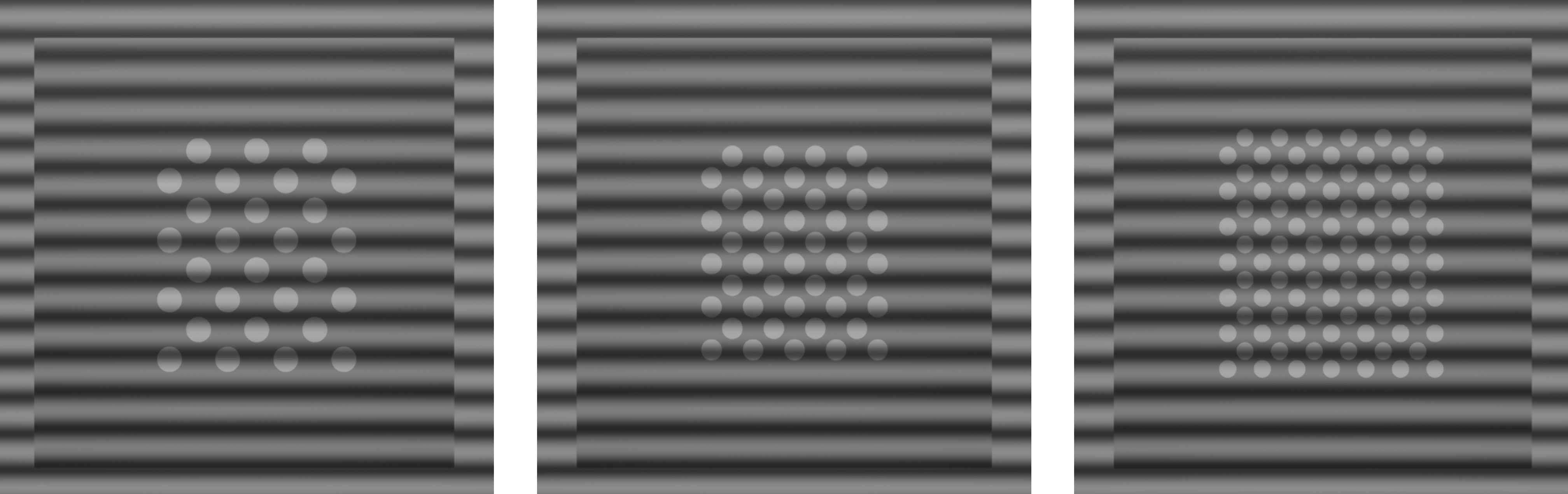}
    \caption{Example generated circular calibration boards: 4$\times$7 (left),
    5$\times$9 (middle), and 7$\times$13 (right).}
    \label{fig:gencalib}
\end{figure}

\noindent These scaled values, $D_{\rm circle}^{\rm sim}$ and $D_{\rm centers}^{\rm sim}$,
are passed directly to the calibration algorithm as the known board geometry.
Example calibration board configurations are shown in Figure~\ref{fig:gencalib}.

\subsection{Virtual Calibration Process}

\begin{figure}[ht]
    \includegraphics[width=1.0\linewidth]{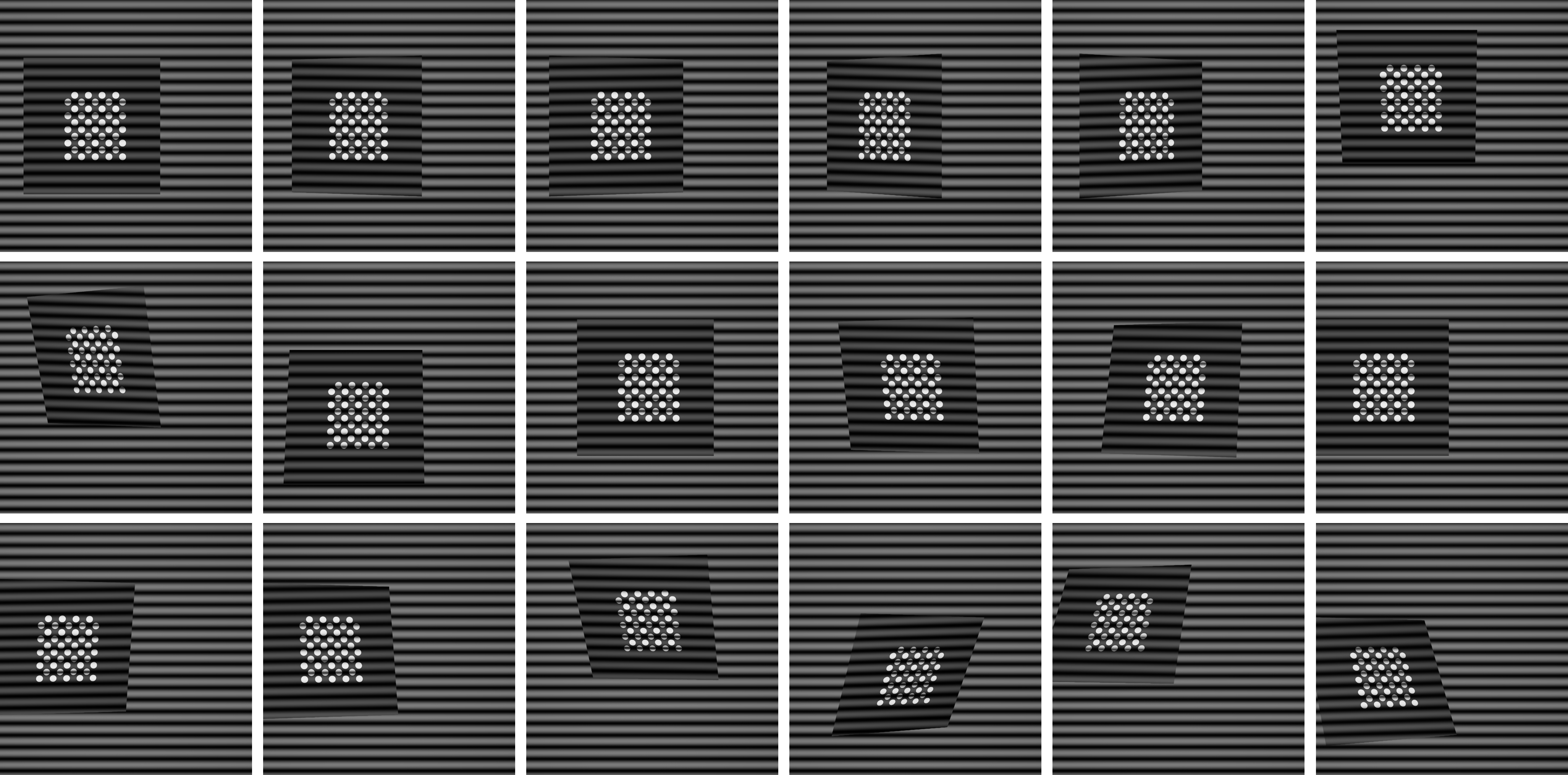}
    \caption{Automatic repositioning of the calibration board across 18 distinct poses for
    virtual camera-projector calibration.}
    \label{fig:calibposes}
\end{figure}

We acquire 18 unique calibration board poses with vertical/lateral translations and angular tilts to provide comprehensive coverage across the camera-projector field of view and working volume (Fig.~\ref{fig:calibposes}).
The calibration follows
Algorithm~\ref{alg:virtual_calib}.
The resulting stereo reprojection error (0.056 px) and projector error (0.049 px)
demonstrate sub-pixel accuracy.

\begin{algorithm}[!t]
\caption{Virtual FPP System Calibration}\label{alg:virtual_calib}
\begin{algorithmic}[1]
\REQUIRE Fringe patterns $\{F_1,\ldots,F_N\}$, Poses $\{P_1,\ldots,P_{18}\}$
\ENSURE Calibrated $M^c$, $M^p$
\STATE Initialize simulation; load fringe textures
\FOR{$i = 1$ \textbf{to} $18$}
    \STATE Position board at pose $P_i$
    \FOR{$j = 1$ \textbf{to} $N$}
        \STATE Project $F_j$; capture frame $I_{i,j}$
    \ENDFOR
    \STATE Extract circle features; compute phase maps (Eq.~\ref{pwrap},\ref{puwrap})
\ENDFOR
\STATE Camera calibration $\rightarrow$ stereo calibration $\rightarrow$ compute errors
\IF{errors $<$ threshold} \RETURN $M^c, M^p$
\ELSE
    \STATE Refine and repeat
\ENDIF
\end{algorithmic}
\end{algorithm}

\section{Experimental Validation and Results}
\label{validation}

All experiments were conducted on an HP OMEN MAX 16 laptop running Windows 11, with an Intel Core Ultra 7 155H CPU, 32 GB DDR5 RAM, and an NVIDIA GeForce RTX 5080 Laptop GPU with 16 GB VRAM (driver 591.44, CUDA 13.1). Simulations used NVIDIA Isaac Sim 4.1.0.

\subsection{Reconstruction Accuracy Validation}

\begin{figure*}[!htbp]
    \centering
    \subfloat[Fringe capture]{%
        \includegraphics[width=0.21\textwidth]{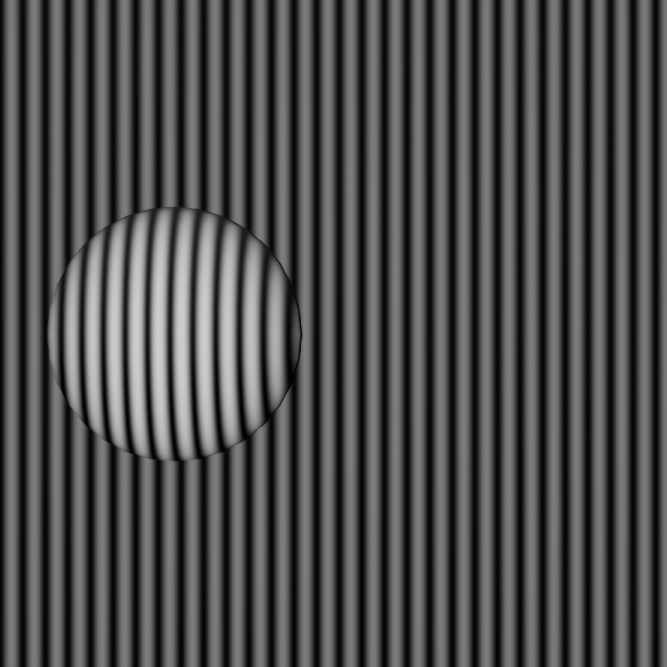}}
    \hfill
    \subfloat[Mesh visualization]{%
        \includegraphics[width=0.21\textwidth]{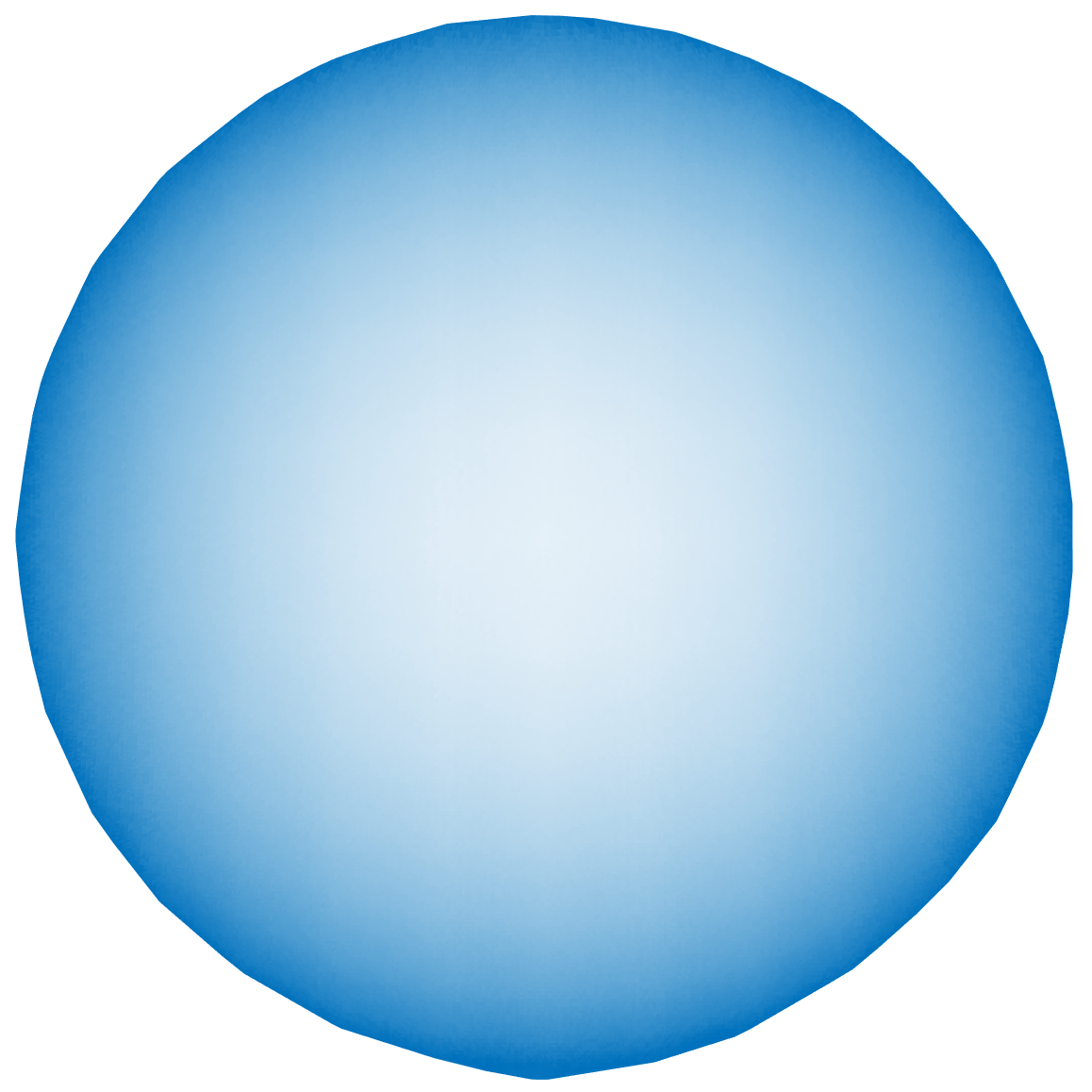}} 
    \hfill
    \subfloat[Sphere fitting]{%
        \includegraphics[width=0.27\textwidth]{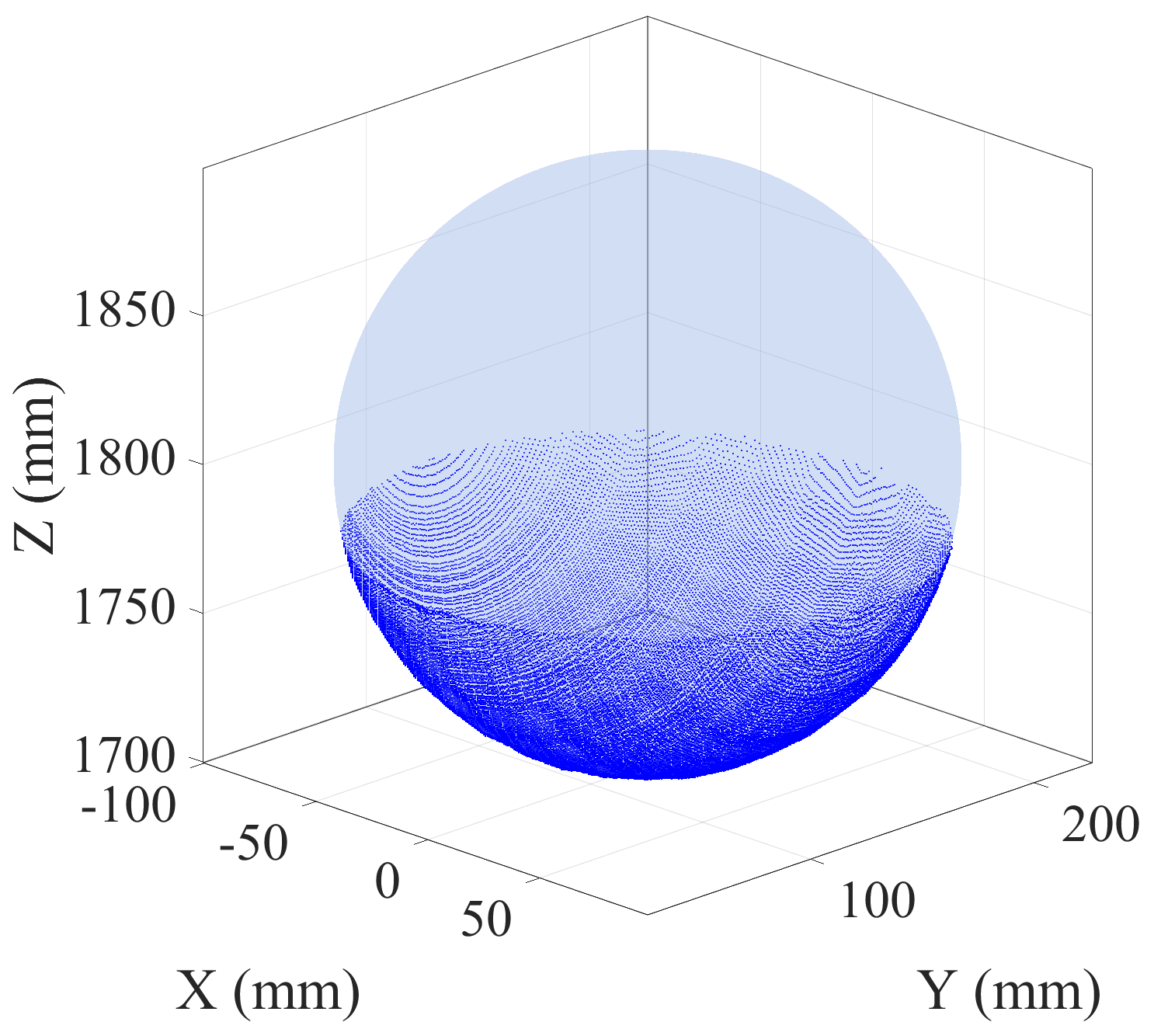}}
    \hfill
    \subfloat[Radial error distribution]{%
        \includegraphics[width=0.27\textwidth]{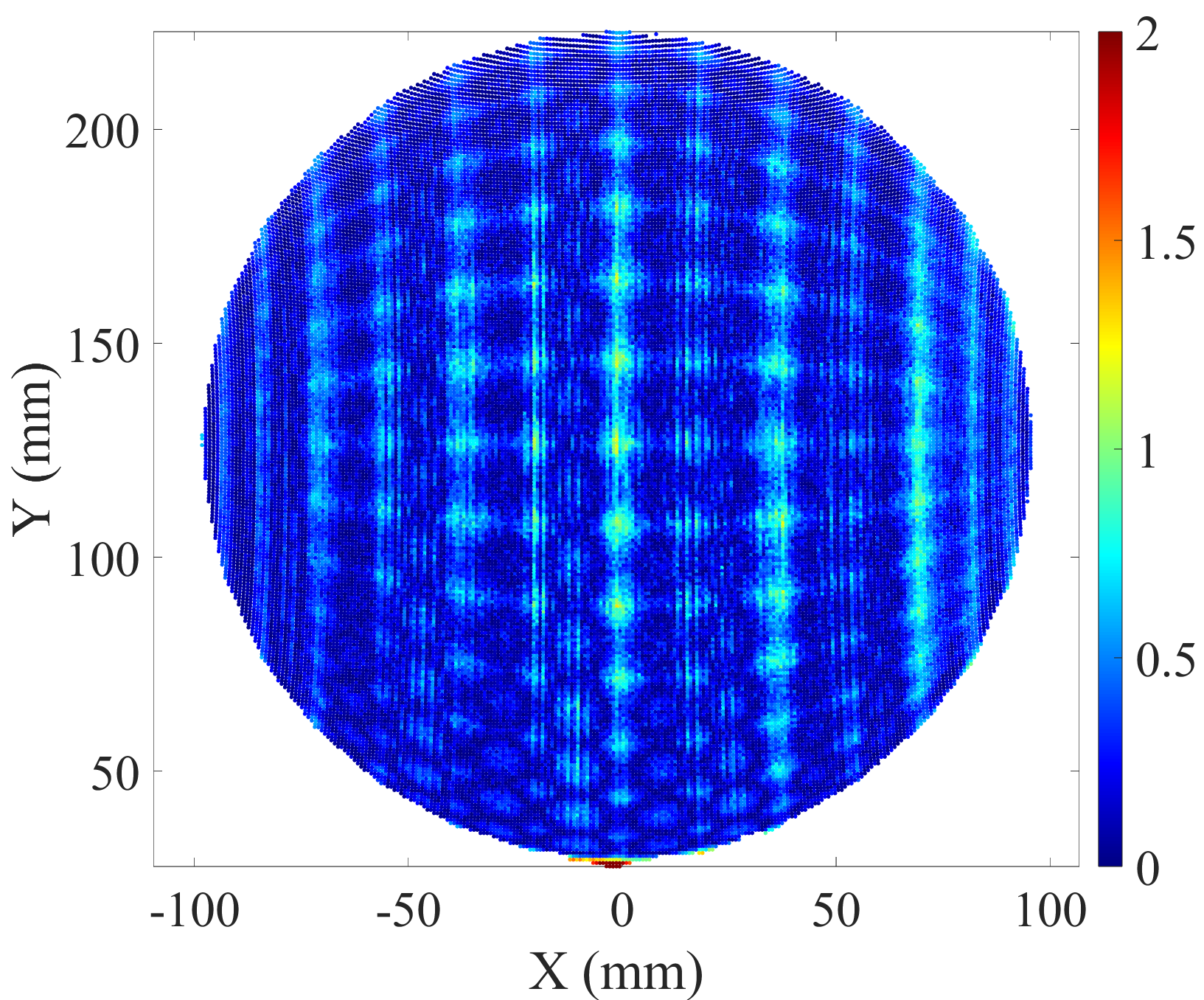}}
    \caption{Sphere reconstruction (radius 100 mm): (a) fringe pattern, (b) reconstructed
    mesh, (c) MSAC sphere fit overlay, (d) radial error distribution (0.661 mm).}
    \label{fig:sphere_fit}
\end{figure*}

We performed 3D reconstruction of a standard sphere mesh (radius 100 mm) using the
virtually calibrated parameters.
We utilized
MATLAB's \texttt{pcfitsphere} (MSAC) to fit the sphere to obtain its fitted radius and inlier percentage~\cite{torr2000mlesac}.
Results are shown in Figure~\ref{fig:sphere_fit}.
The fitted radius was 99.3391 mm (98.6\% inliers: 48\,298/48\,975), giving:
\begin{equation}\label{eqn:radial_err}
    R_{\text{abs}} = |99.339 - 100.0| = 0.661\text{ mm}\quad(0.661\%)
\end{equation}
This sub-millimeter accuracy is comparable to physical FPP
systems~\cite{lv2023modeling}, validating the virtual calibration methodology.

\subsection{Digital Twin Validation} \label{Digital Twin Validation}

To replicate a physical FPP system in simulation, camera intrinsic parameters from
stereo-calibration are transferred to Isaac Sim's camera model via~\cite{cam_docs}:
\begin{align}
f = \tfrac{f_x+f_y}{2}\cdot s_{\text{mm}}, \quad
A_h = W\cdot s_{\text{mm}}, \quad
A_v = H\cdot s_{\text{mm}},
\end{align}
where $s_{\text{mm}}$ is the physical pixel size and $[W,H]$ is image resolution.

\begin{figure}
    \includegraphics[width=0.5\textwidth]{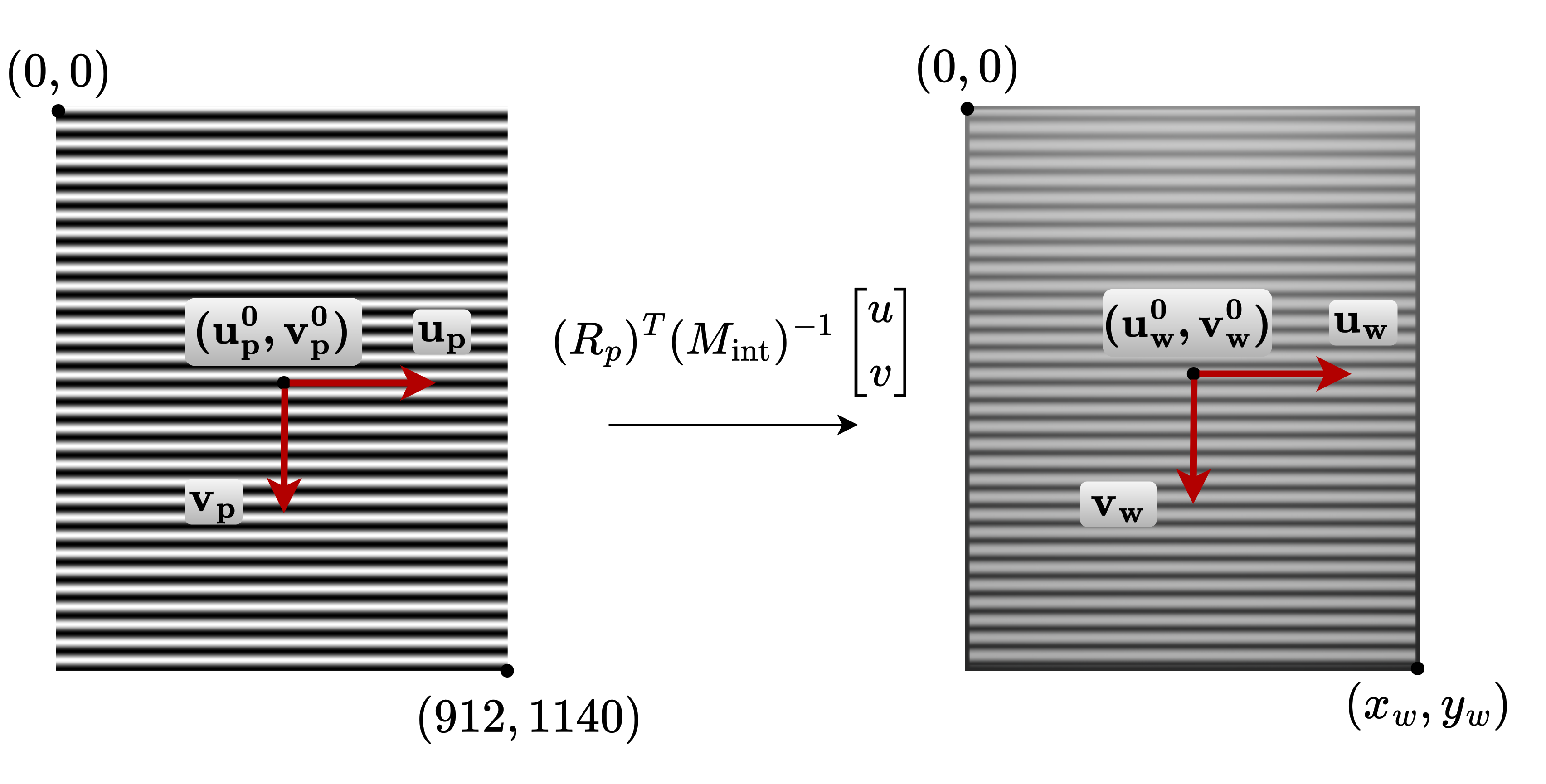}
    \caption{Coordinate transformation from projector image space to world coordinates for
    determining projected pattern metric dimensions.}
    \label{fig:p2w}
\end{figure}

\begin{figure}[htbp]
    \centering
    {\includegraphics[width=0.5\textwidth]{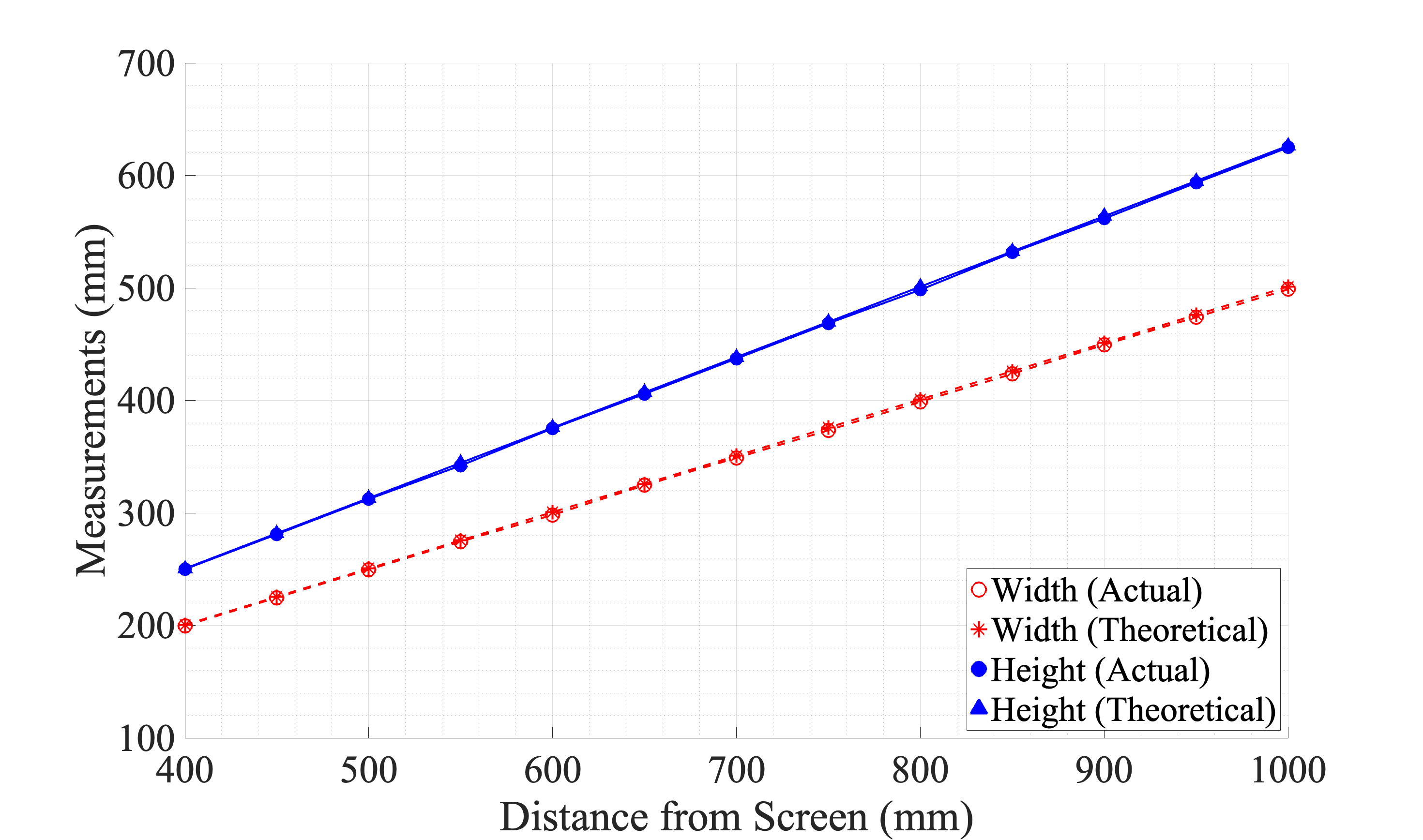}\caption*{(a)}}
    {\includegraphics[width=0.5\textwidth]{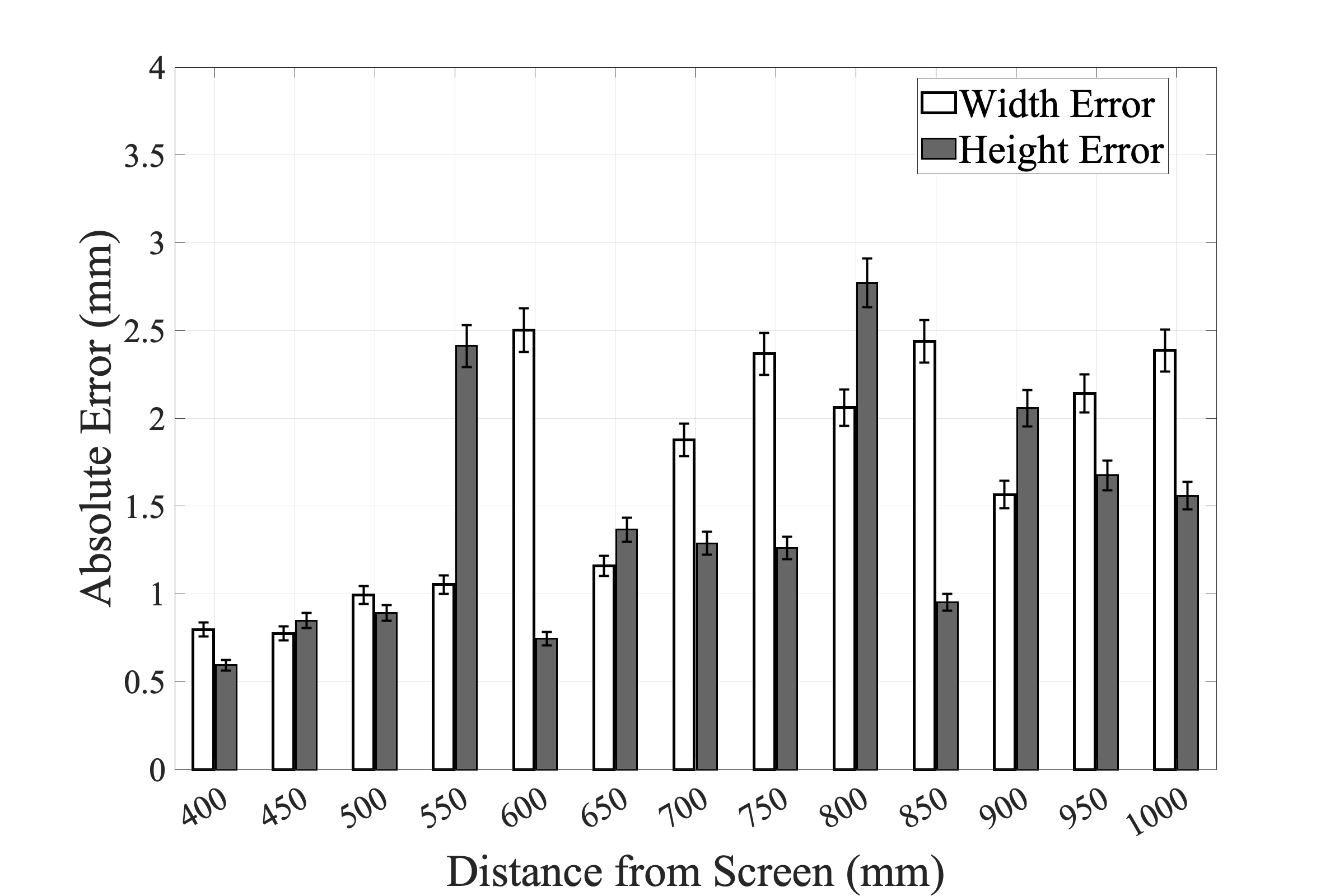}\caption*{(b)}}
    \caption{(a) Measured (single sample per distance via Isaac Sim's built-in Measurement Tool) and theoretical projected fringe width/height versus distance; (b) absolute error between measured and theoretical values at each distance. Error bars indicate the $\pm5\%$ measurement-tool placement uncertainty arising from manual cursor positioning on the projected pattern edges. The mean absolute error across all distances is 1.70 mm (width) and 1.42 mm (height).}
    \label{fig:linearity_viz}
\end{figure}

Isaac Sim does not provide a direct interface for projector calibration matrices; projectors
are modeled as rectangular light sources whose physical dimensions must be specified.
Direct physical measurement is impractical because projected pattern dimensions vary with
working distance and introduce systematic errors.
We instead derive metric projector dimensions analytically by back-projecting projector pixels through the inverse camera formulation of the projector~\cite{martynov2011projector,hartley2003pinhole}. A projector pixel $[u, v]^T$ defines a ray originating at the projector's optical center $\mathbf{o}_p = -R_p^T t_p$ and propagating along the direction
\begin{equation}\label{eq:rayworld}
\mathbf{d}_w = R_p^{-1} M_{\text{int}}^{-1}\begin{bmatrix} u \\ v \\ 1 \end{bmatrix}
\end{equation}
in world coordinates, where $M_{\text{int}}$ and $M_{\text{ext}} = [R_p\,|\,t_p]$
are the projector intrinsic and extrinsic matrices. The intrinsic inverse is
evaluated analytically, and the rotation block $R_p$ is inverted directly, as it
is orthonormal and therefore non-singular, with $R_p^{-1} = R_p^T$. The
translation vector $t_p$ is not used in the back-projection: it only locates the
projector's optical center $\mathbf{o}_p$, whereas the ray direction
$\mathbf{d}_w$ is determined solely by $R_p$ and $M_{\text{int}}$ and is therefore
invariant to $t_p$~\cite{hartley2003pinhole}. Because a 2D pixel back-projects to
a one-parameter family of 3D points $\mathbf{o}_p + \lambda\,\mathbf{d}_w$ along
this ray, recovering a unique world point requires an additional depth
constraint. Enforcing the pinhole projector constraint that projected pattern
dimensions scale linearly with working distance $Z$~\cite{hartley2003pinhole}
closes the system, giving the closed-form solution:
\begin{equation}\label{eq:p2wfin}
\begin{bmatrix} x_w \\ y_w \\ 1 \end{bmatrix}
= Z\!\left(\left(R_p\right)^{T}\!\left(M_{\text{int}}\right)^{-1}
\begin{bmatrix} u \\ v \\ 1 \end{bmatrix}\right)
\end{equation}
The parenthesized term is the back-projected ray direction $\mathbf{d}_w$,
and the working distance $Z$ fixes the plane at which the projector's ray bundle
is sampled.
The projected image width and height are obtained by evaluating Eq.~\ref{eq:p2wfin} at
corner pixels $[0,0]$ and $[912,1140]$ (Fig.~\ref{fig:p2w}).
This rectangular light source combined with the inverse camera model is mathematically
equivalent to a pinhole projector: projected pattern dimensions scale linearly with
distance, validated empirically from 400--1000 mm (MAE $<$ 2 mm;
Fig.~\ref{fig:linearity_viz}).

\textbf{Inverse model validation:}
Applied to virtual system calibration matrices (board at 1 m; stereo error: 0.066 px,
projector error: 0.058 px), Eq.~\ref{eq:p2wfin} yields $[x_w,y_w]=[626.3, 501.1]$ mm,
closely matching the Table~\ref{tab:cam_proj_params} projector dimensions
(625$\times$500 mm).
Figure~\ref{fig:linearity_viz} confirms the expected linear perspective scaling across
400--1000 mm (MAE: 1.70 mm width, 1.42 mm height).

\begin{figure}[ht]
\centering
    \includegraphics[width=0.95\linewidth]{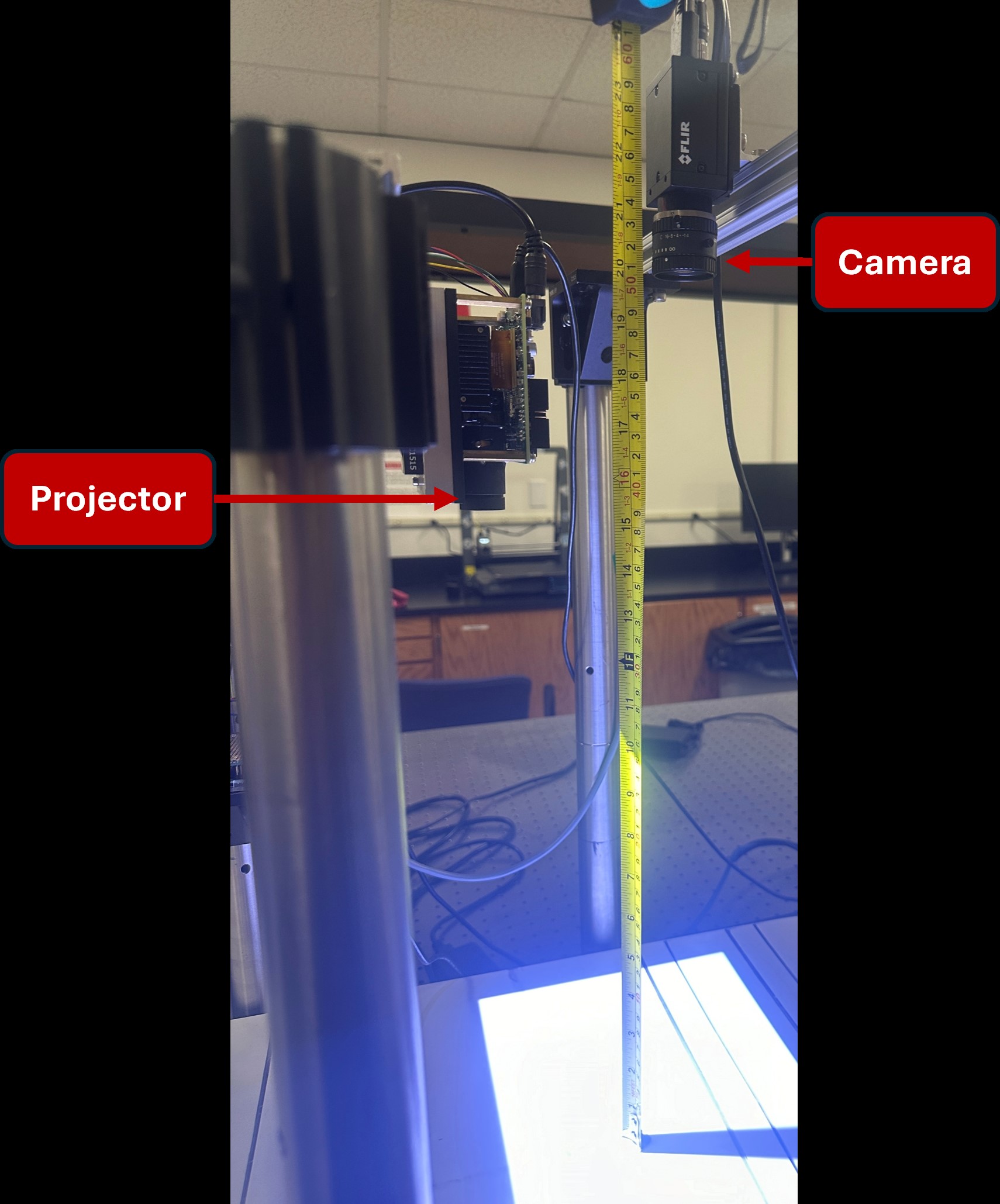}
    \caption{Physical FPP system used for digital twin validation; measuring tape shows the
    400 mm screen distance.}
    \label{fig:real-system}
\end{figure}

\begin{figure}[htbp]
    \centering
    \includegraphics[width=0.5\textwidth]{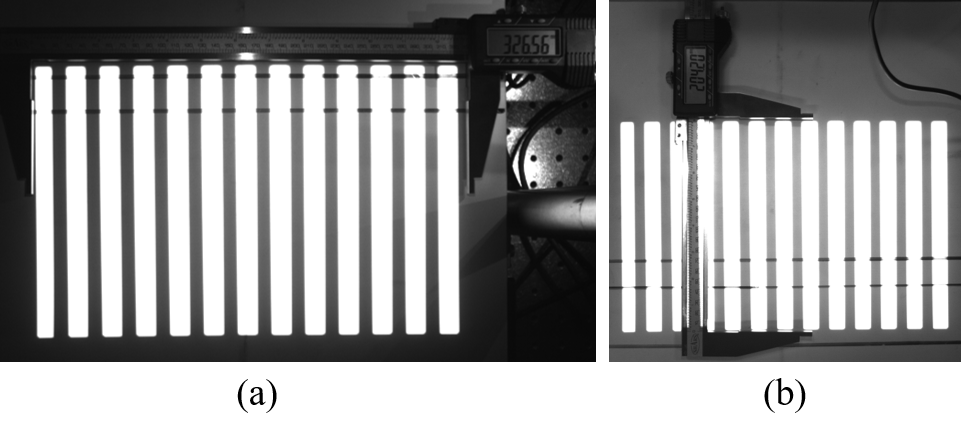}
    \caption{Real-world fringe dimensions at 0.4 m: (a) Height = 326.56 mm;
    (b) Width = 204.20 mm.}
    \label{fig:real_measurements}
\end{figure}

\textbf{Real-world digital twin:}
Applying Eq.~\ref{eq:p2wfin} to physical system calibration matrices yields
$[x_w,y_w]=[202.7, 323.3]$ mm.
Comparative fringe dimension measurements at 0.4 m confirm real-to-sim agreement
(Fig.~\ref{fig:real_measurements},\ref{fig:real-system}).
We found that setting projector attributes to the values from Eq.~\ref{eq:p2wfin}
at $Z=1$ m achieves accurate dimensional correspondence, consistent with Isaac Sim's
\texttt{RectLight} primitive behavior, which we attribute to 1 m being the
implementation's reference projection distance.

\subsection{3D Reconstruction}\label{3D Reconstruction}

\begin{figure*}[!htbp]
    \centering
    {\includegraphics[width=0.3\textwidth]{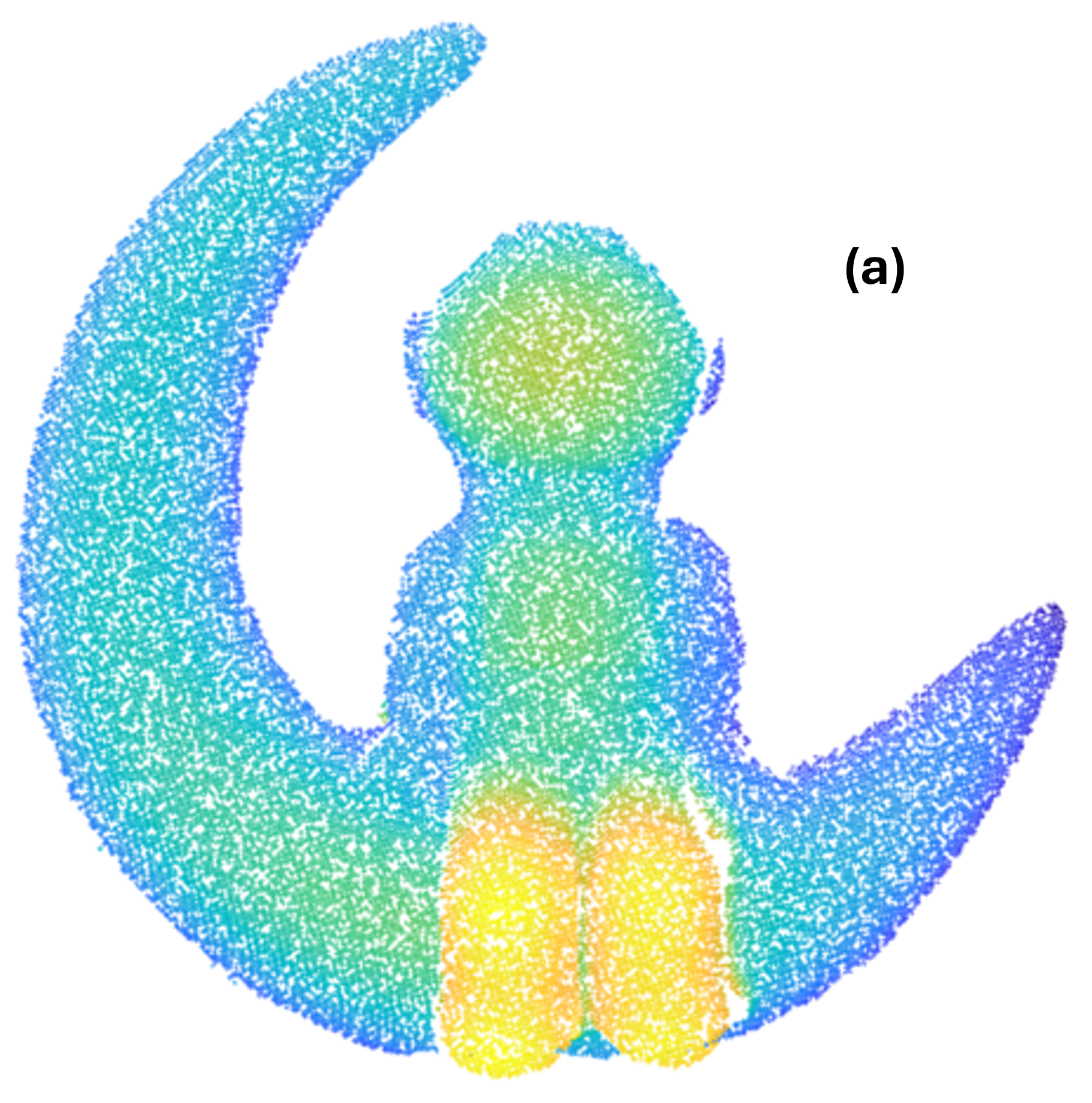}}
    {\includegraphics[width=0.4\textwidth]{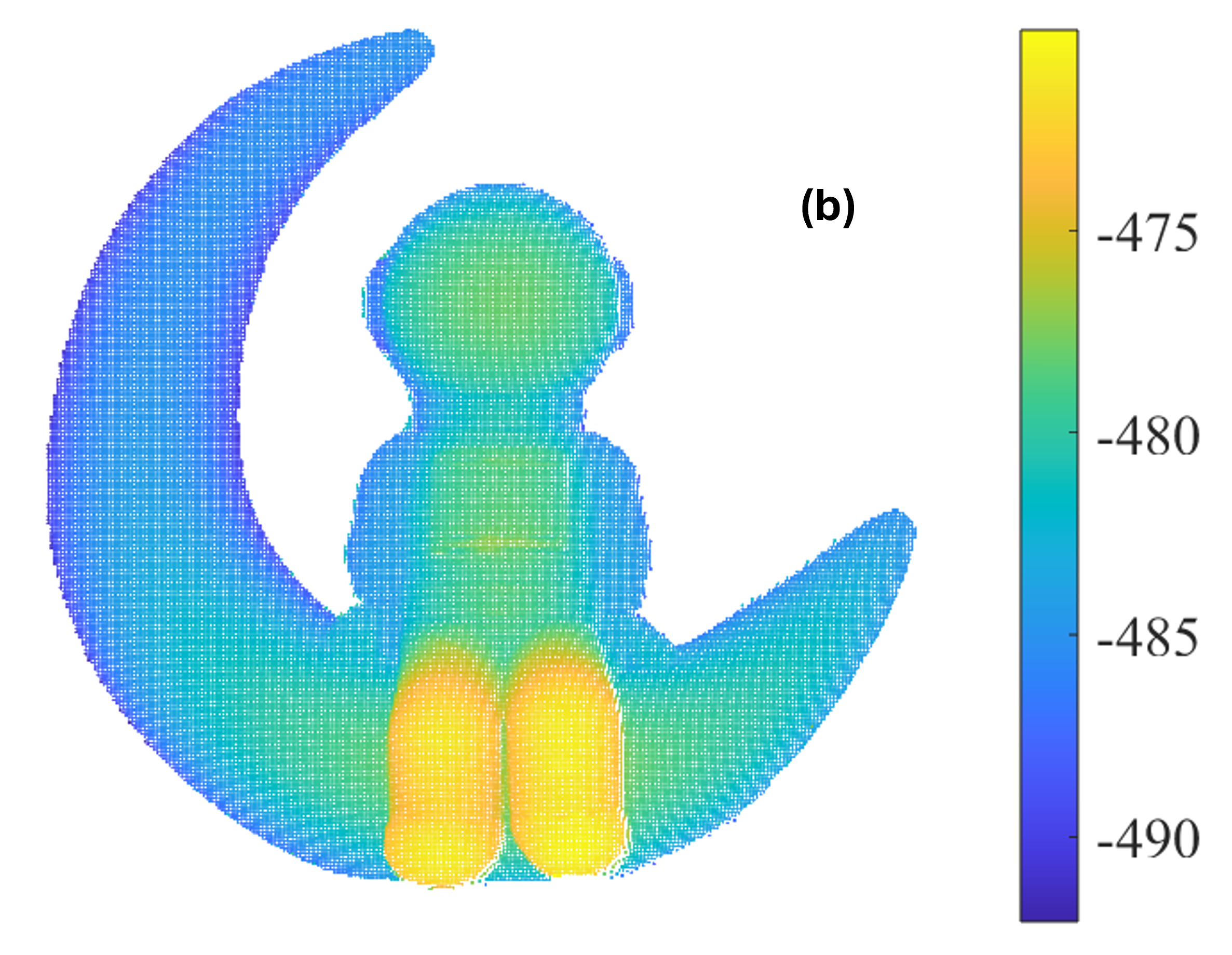}}
    \hfill
    {\includegraphics[width=0.7\textwidth]{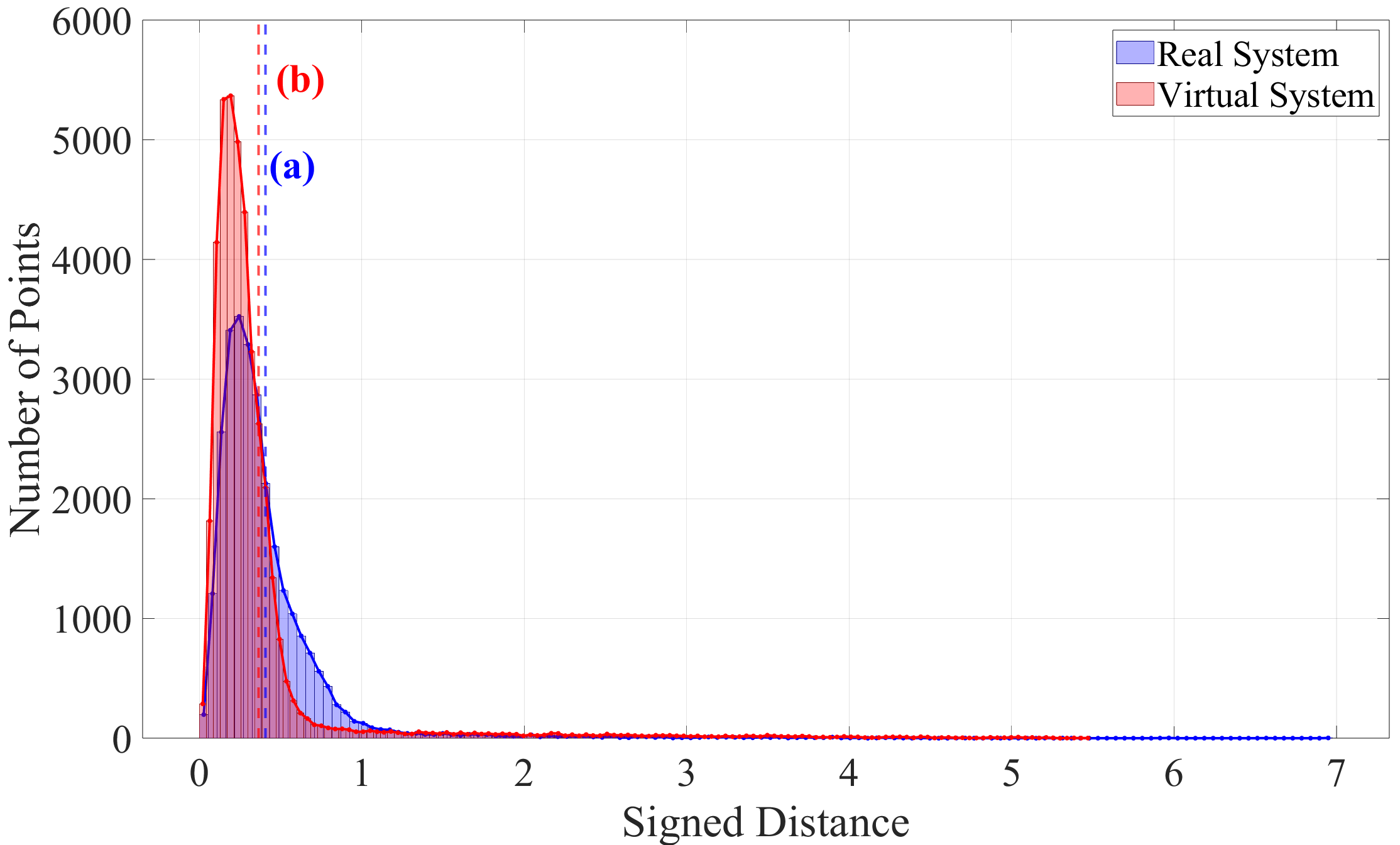}}
        \caption*{(c)}
    \caption{3D reconstructions of astronaut figurine: (a) physical and (b) virtual outputs,
    both evaluated against the STL model via C2M distance. (c) C2M histogram: virtual (red)
    mean = 0.37 mm; real (blue) mean = 0.41 mm (Table~\ref{tab:stl_errors}).}
    \label{fig:reconstructions_comparison}
\end{figure*}

\begin{figure}
    \centering
    \includegraphics[width=0.85\columnwidth]{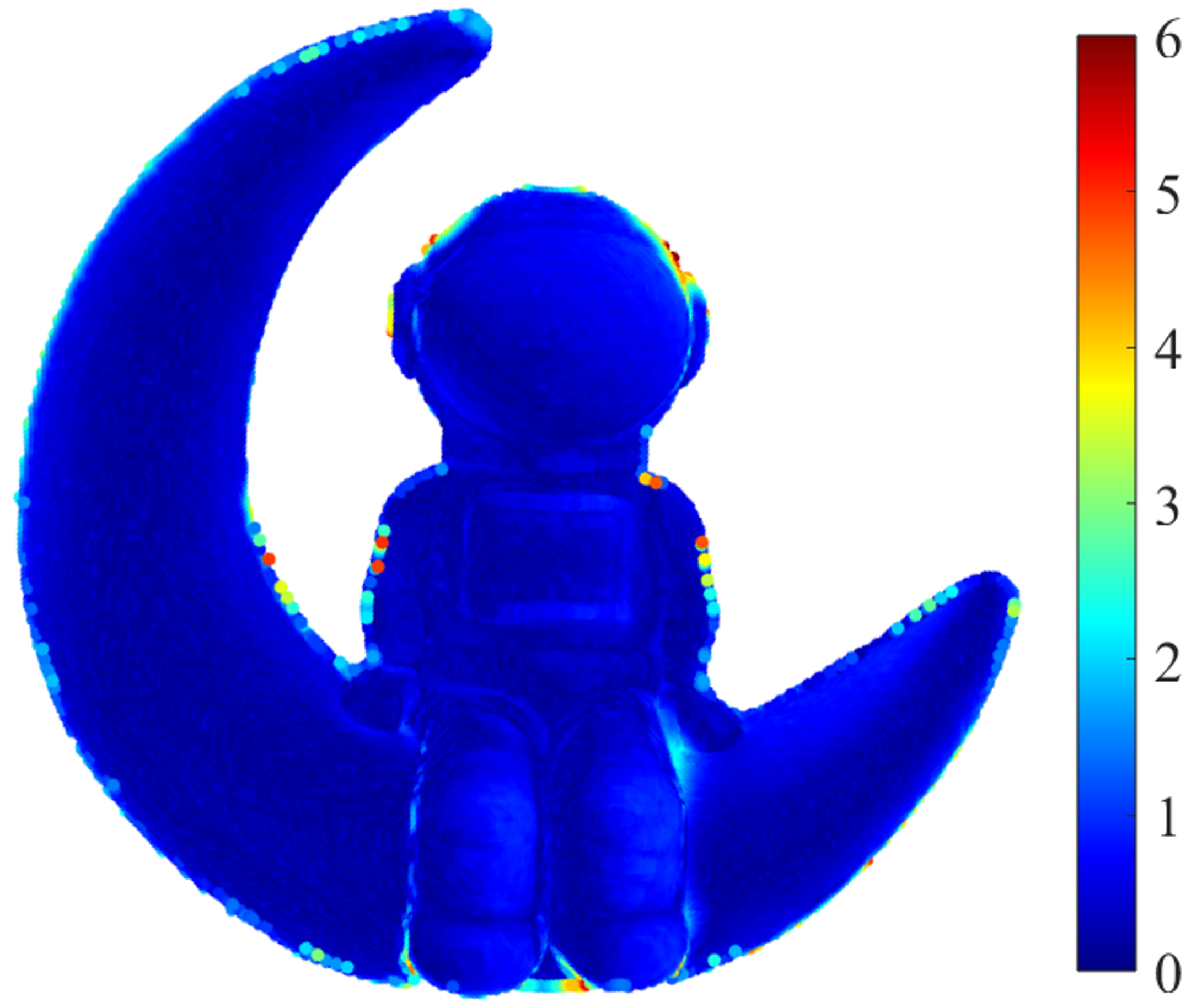}
    \caption{Inter-system Euclidean distance between virtual and real point clouds: MAE = 0.436 mm and RMSE = 0.578 mm, confirming digital twin fidelity.}
    \label{fig:dtvalidation}
\end{figure}

With both systems calibrated, we scanned a 3D-printed astronaut figurine using the digital
twin and compared virtual and physical reconstructions against the original STL model via
Cloud-to-Mesh (C2M) distance analysis with ICP
registration~\cite{lakshman2025characterizing}.
Figure~\ref{fig:reconstructions_comparison} and Table~\ref{tab:stl_errors} show the
majority of points for both systems fall within 0--1 mm.
The inter-system Euclidean distance between point clouds (Fig.~\ref{fig:dtvalidation})
yields MAE = 0.436 mm and RMSE = 0.578 mm, validating digital twin fidelity.

\begin{table}
\caption{C2M reconstruction accuracy against reference STL.}
\label{tab:stl_errors}
\centering
\begin{tabular}{c|c|c}
\hline\hline
\textbf{System} & \textbf{MAE} & \textbf{RMSE} \\
\hline
Real    & 0.41 mm & 0.56 mm \\
Virtual & 0.37 mm & 0.65 mm \\
\hline\hline
\end{tabular}
\end{table}

\section{Discussion}
\label{discussion}

VIRTUS-FPP offers three principal advantages over physical FPP systems.
(1)~\textbf{Sensor design flexibility}: programmatic control of all optical parameters enables
rapid virtual prototyping before committing to hardware.
(2)~\textbf{Controlled testing}: environmental conditions (lighting, material properties, geometry) can be precisely
controlled at scale, which is impractical with physical rigs.
(3)~\textbf{Perfect ground truth}: exact geometric reference enables quantitative error analysis
not achievable with physical measurement instruments.

Known limitations include: (1)~\textbf{Computational requirements}: a minimum RTX 3070
with 8 GB VRAM is required; lower-spec hardware reduces capture throughput significantly.
(2)~\textbf{Rendering noise}: sampled direct lighting, a renderer mode that stochastically samples a subset of light sources per shading point rather than evaluating contributions from all emitters, must be disabled in the real-time ray-tracing settings. This mode is optimized for scenes containing many lights, but its stochastic sampling introduces high-variance noise on the high-frequency sinusoidal fringe textures, producing pixel-level phase artifacts that degrade reconstruction accuracy. Deterministic evaluation through the projected texture is required for accurate fringe rendering.
(3)~\textbf{Simulation scope}: the framework models radiometric light transport at the
level of geometric optics and surface reflectance via RTX ray tracing.
Wave-optical phenomena (coherence, diffraction, interference) and device-specific
nonlinearities (projector gamma response, lens distortion, sensor noise) are not explicitly
modeled, which may limit direct transfer of algorithms sensitive to these effects.
Extending the framework with calibrated gamma and lens distortion models is a planned future
direction.

Measurement uncertainty arises from calibration residuals (0.056 px stereo; 0.049 px
projector), phase calculation, and triangulation geometry.
The 0.37 mm digital twin MAE provides a practical bound on the combined effect, consistent
with physical FPP precision~\cite{lv2023modeling}.

\begin{figure}[ht]
    \centering
    \includegraphics[width=1.0\linewidth]{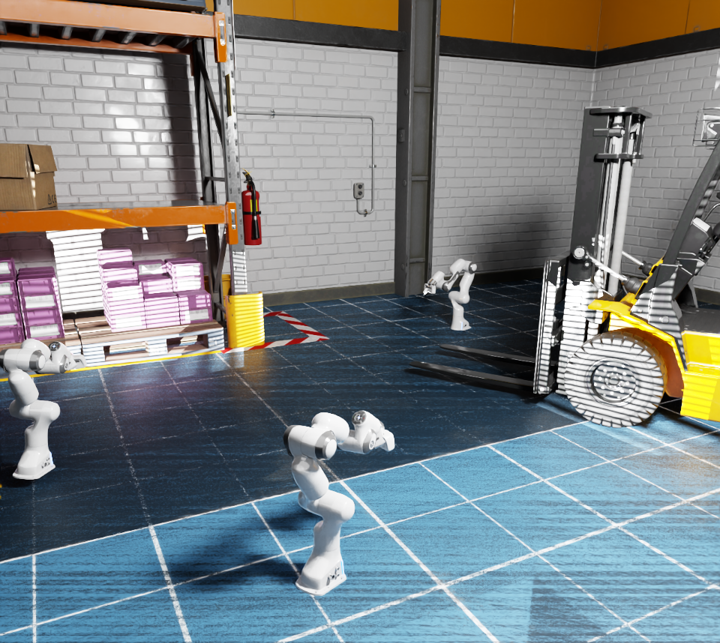}
    \caption{Franka manipulators each carrying the simulated FPP camera-projector pair, scanning distinct objects in Isaac Sim's \texttt{warehouse\_with\_forklifts} environment to illustrate native integration of the optical metrology pipeline into robotic simulation.}
    \label{fig:robot_scene}
\end{figure}

\textbf{Robotics integration.} Beyond the optical-metrology pipeline, Isaac Sim's robotics-native architecture enables seamless integration of VIRTUS-FPP into robotic tasks. Figure~\ref{fig:robot_scene} shows an example deployment in which multiple Franka manipulators, each equipped with the modeled FPP sensor, simultaneously scan distinct objects within Isaac Sim's pre-defined \texttt{warehouse\_with\_forklifts} environment. This integration unlocks application classes that are difficult to prototype on physical hardware, including large-scale autonomous mobile 3D scanning, eye-in-hand sensor placement optimization, coordinated multi-robot inspection, and synthetic data generation for learning-based grasping policies that depend on accurate surface geometry.

{\textbf{Machine learning applications.} VIRTUS-FPP also supports machine learning workflows that require large, ground-truth-annotated training sets. We have applied the framework to generate photorealistic synthetic data for comprehensive benchmarking of learning-based methods for single-shot long-range fringe analysis~\cite{lakshman2026benchmarking}, a regime where annotated physical datasets are scarce. This downstream demonstration confirms VIRTUS-FPP's effectiveness as a scalable source of training data for ML-driven optical metrology research.

\section{Conclusion and Future Work}
\label{conclusion}

We presented VIRTUS-FPP, the first end-to-end virtual FPP sensor modeling framework
built on NVIDIA Isaac Sim, providing projective geometric calibration, RTX ray-traced
light transport, and calibrated metric consistency across the complete imaging pipeline.
Key contributions: (1)~a complete virtual camera-projector sensor system without physical
hardware dependence; (2)~virtual calibration achieving sub-pixel accuracy (0.056 px stereo,
0.049 px projector); (3)~sub-millimeter reconstruction accuracy (0.661 mm on a 100 mm
sphere); and (4)~a validated digital twin of a real FPP system achieving 0.436 mm inter-system MAE via
the inverse camera model.
VIRTUS-FPP reduces FPP sensor development complexity and enables large-scale synthetic data
generation with exact ground truth for machine learning applications.
Future work includes extending the simulation scope with calibrated projector gamma and lens
distortion models, multi-view FPP configurations, additional validation objects with complex
geometries (cylinders, freeform surfaces), robotic manipulation integration, and domain
randomization experiments for sim-to-real transfer.

\section*{Code Availability}
The VIRTUS-FPP source code is available at \url{https://github.com/oadamharoon/virtus-fpp}.

\bibliographystyle{ieeetr}
\bibliography{vsmfpp}

\vfill

\end{document}